\documentclass[12pt,a4paper]{article}
\usepackage{amssymb,amsmath,bm}
\usepackage{graphicx}

\def\simge{\mathrel{\rlap{\raise 0.511ex \hbox{$>$}}{\lower 0.511ex \hbox{$\sim$}}}}
\def\simle{\mathrel{\rlap{\raise 0.511ex \hbox{$<$}}{\lower 0.511ex \hbox{$\sim$}}}}
\def\slash#1{\setbox0=\hbox{$#1$}\dimen0=\wd0
     \setbox1=\hbox{/} \dimen1=\wd1 \ifdim\dimen0>\dimen1
     \rlap{\hbox to \dimen0{\hfil/\hfil}} #1                        \else
     \rlap{\hbox to \dimen1{\hfil$#1$\hfil}}
     /   \fi}

\def\slash#1{\mbox{$\not \!\! #1$}}

\def\lvec#1{\setbox0=\hbox{$#1$}
    \setbox1=\hbox{$\scriptstyle\leftarrow$}
    #1\kern-\wd0\smash{
    \raise\ht0\hbox{$\raise1pt\hbox{$\scriptstyle\leftarrow$}$}}
    \kern-\wd1\kern\wd0}
\def\rvec#1{\setbox0=\hbox{$#1$}
    \setbox1=\hbox{$\scriptstyle\rightarrow$}
    #1\kern-\wd0\smash{
    \raise\ht0\hbox{$\raise1pt\hbox{$\scriptstyle\rightarrow$}$}}
    \kern-\wd1\kern\wd0}

\def\diracstar#1#2{
    \setbox0=\hbox{$\gamma$}\setbox1=\hbox{$\gamma_{#1}$}
    \gamma_{#1}\kern-\wd1\kern\wd0
    \smash{\raise4.5pt\hbox{$\scriptstyle#2$}}}

\newcommand{\lsim}{
\mathrel{\hbox{\rlap{\hbox{\lower4pt\hbox{$\sim$}}}\hbox{$<$}}}}

\newcommand{\gsim}{
\mathrel{\hbox{\rlap{\hbox{\lower4pt\hbox{$\sim$}}}\hbox{$>$}}}}

\newcommand{\gev}{\, {\rm GeV}}
\newcommand{\mev}{\, {\rm MeV}}

\newcommand{\be}{\begin{equation}}
\newcommand{\ee}{\end{equation}}
\newcommand{\bea}{\begin{eqnarray}}
\newcommand{\eea}{\end{eqnarray}}
\newcommand{\nn}{\nonumber}

\begin{document}

\centerline{\huge Leptonic decay constants $f_K$, $f_D$ and $f_{D_s}$} 
\vspace{0.5cm}
\centerline{\huge with $N_f = 2+1+1$ twisted-mass lattice QCD}

\vspace{1.5cm}

\centerline{\large N.~Carrasco$^{(a)}$, P.~Dimopoulos$^{(b,c)}$, R.~Frezzotti$^{(c,d)}$, P.~Lami$^{(e,a)}$,}  
\vspace{0.25cm}
\centerline{\large V.~Lubicz$^{(e,a)}$, F.~Nazzaro$^{(c)}$, E.~Picca$^{(e,a)}$, L.~Riggio$^{(e,a)}$,}  
\vspace{0.25cm}
\centerline{\large G.C.~Rossi$^{(b,c,d)}$, F.~Sanfilippo$^{(f)}$, S.~Simula$^{(a)}$, C.~Tarantino$^{(e,a)}$}

\vspace{1cm}

\centerline{\includegraphics[draft=false]{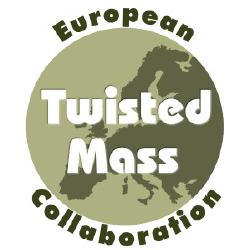}}

\vspace{1cm}

\centerline{\it $^{(a)}$ INFN, Sezione di Roma Tre}
\centerline{\it Via della Vasca Navale 84, I-00146 Rome, Italy}
\vskip 2 true mm
\centerline{\it $^{(b)}$ Centro Fermi - Museo Storico della Fisica e Centro Studi e Ricerche Enrico Fermi}
\centerline{\it Compendio del Viminale, Piazza del Viminale 1 I–00184 Rome, Italy }
\vskip 2 true mm
\centerline{\it $^{(c)}$ Dipartimento di Fisica, Universit\`a di Roma ``Tor Vergata''}
\centerline{\it Via della Ricerca Scientifica 1, I-00133 Rome, Italy}
\vskip 2 true mm
\centerline{\it $^{(d)}$ INFN, Sezione di ``Tor Vergata"}
\centerline{\it Via della Ricerca Scientifica 1, I-00133 Rome, Italy}
\vskip 2 true mm
\centerline{\it $^{(e)}$ Dipartimento di Matematica e Fisica, Universit\`a degli Studi Roma Tre}
\centerline{\it Via della Vasca Navale 84, I-00146 Rome, Italy}
\vskip 2 true mm
\centerline{\it $^{(f)}$ School of Physics and Astronomy, University of Southampton,}
\centerline{\it SO17 1BJ Southampton, United Kindgdom}

\newpage

\begin{abstract}
We present a lattice QCD calculation of the pseudoscalar decay constants $f_K$, $f_D$ and $f_{D_s}$ performed using the gauge configurations produced by the European Twisted Mass Collaboration with $N_f = 2 + 1 + 1$ dynamical quarks, which include in the sea, besides two light mass degenerate quarks, also the strange and charm quarks with masses close to their values in the real world. 
The simulations are based on a unitary setup for the two light mass-degenerate quarks and on a mixed action approach for the strange and charm quarks.
We use data simulated at three different values of the lattice spacing in the range $0.06 \div 0.09$ fm and at pion masses in the range $210 \div 450$ MeV. 
Our main results are: $f_{K^+} / f_{\pi^+} = 1.184 (16)$, $f_{K^+} = 154.4 (2.0) \mev$, which incorporate the leading strong isospin breaking correction due to the up- and down-quark mass difference, and $f_K = 155.0 (1.9) \mev$, $f_D = 207.4 (3.8) \mev$, $f_{D_s} = 247.2 (4.1) \mev$, $f_{D_s} / f_D = 1.192 (22)$ and $(f_{D_s} / f_D) / (f_K / f_\pi) = 1.003 (14)$ obtained in the isospin symmetric limit of QCD.
Combined with the experimental measurements of the leptonic decay rates of kaon, pion, $D$- and $D_s$-mesons our results lead to the following determination of the CKM matrix elements: $|V_{us}| = 0.2269 (29)$, $|V_{cd}| = 0.2221 (67)$ and $|V_{cs}| = 1.014 (24)$.
Using the latest value of $|V_{ud}|$ from superallowed nuclear $\beta$ decays the unitarity of the first row of the CKM matrix is fulfilled at the permille level.
\end{abstract}

\newpage

\section{Introduction}
\label{sec:intro}

The leptonic decay constants of light and heavy pseudoscalar (PS) mesons are the crucial hadronic ingredients necessary for obtaining precise information on the Cabibbo-Kobayashi-Maskawa (CKM) quark-mixing matrix elements \cite{CKM} within the Standard Model.
Together with the experimental data for the ratio of decay rates $\Gamma(K^+ \to \mu^+ \nu) / \Gamma(\pi^+ \to \mu^+ \nu)$ and with the average value of the CKM matrix element $|V_{ud}|$ from superallowed nuclear beta decays, the ratio $f_{K^+} / f_{\pi^+}$ allows to determine the CKM matrix element $|V_{us}|$ and to test the unitarity relation of the first-row of the CKM matrix (see Ref.~\cite{FLAG} and references therein).
The combination of the charmed-meson decay constants $f_D$ and $f_{D_s}$ with the experimental measurements of the decay rates for $D(D_s) \to \mu \nu$ and $D(D_s) \to \tau \nu$ enables one to determine the CKM matrix elements $|V_{cd}|$ and $|V_{cs}|$ (see again Ref.~\cite{FLAG} and references therein).

In this paper we present a lattice QCD calculation of the $f_{K^+}$, $f_D$ and $f_{D_s}$ decay constants using the ensembles of gauge configurations produced by the European Twisted Mass (ETM) Collaboration with four flavors of dynamical quarks ($N_f = 2+1+1$), which include in the sea, besides two light mass-degenerate quarks, also the strange and the charm quarks with masses close to their values in the real world \cite{Baron:2010bv, Baron:2010th, Baron:2011sf}.

The gauge ensembles and the simulations used in this work are the same adopted in Ref.~\cite{Carrasco:2014cwa} to determine the up, down, strange and charm quark masses (see Tables 1-3 of Ref.~\cite{Carrasco:2014cwa}), using the experimental value of the pion decay constant, $f_{\pi^+}$, to set the lattice scale\footnote{With respect to Ref.~\cite{Carrasco:2014cwa} the number of gauge configurations adopted for the ensemble D15.48 has been increased to $90$ to improve the statistics.}. 
We employed the Iwasaki action \cite{Iwasaki:1985we} for gluons and the Wilson Twisted Mass Action \cite{Frezzotti:2003xj, Frezzotti:2003ni} for sea quarks. 
In order to avoid the mixing of strange and charm quarks in the valence sector we adopted the non-unitary set up described in Ref.~\cite{Frezzotti:2004wz}, in which the valence strange and charm quarks are regularized as Osterwalder-Seiler (OS) fermions \cite{Osterwalder:1977pc}, while the valence up and down quarks have the same action of the sea.
Working at maximal twist such a setup guarantees an automatic ${\cal{O}}(a)$-improvement \cite{Frezzotti:2003ni, Frezzotti:2004wz}.
We considered three values of the inverse bare lattice coupling $\beta$, which allow for a controlled extrapolation to the continuum limit, and different lattice volumes.
For each gauge ensemble we simulated three values of the valence strange quark mass and six values of the valence heavy quark mass, which are needed for the interpolation in the charm region as well as to extrapolate to the $b$-quark sector for future studies. 
For the light sector we simulated quark masses in the range $3 ~ m_{ud} \lesssim  \mu_\ell \lesssim 12 ~ m_{ud} $, for the strange sector in $0.7 ~ m_s \lesssim  \mu_s \lesssim 1.2 ~ m_s$, while for the charm sector in $0.7 ~ m_c \lesssim  \mu_c \lesssim 2.5 ~ m_c$, where $m_{ud}$, $m_s$ and $m_c$ are the physical values of the average up/down, strange and charm quark masses, respectively, as determined in Ref.~\cite{Carrasco:2014cwa}.
The lattice spacings were found to be $a = \{ 0.0885(36), 0.0815(30), 0.0619(18) \}$ fm at $\beta = \{1.90, 1.95, 2.10\}$ respectively, the lattice volume goes from $\simeq 2$ to $\simeq 3$ fm, and the pion masses, extrapolated to the continuum and infinite volume limits, range from $\simeq 210$ to $ \simeq 450 \mev$ (see Ref.~\cite{Carrasco:2014cwa} for further details).

We present our study of the PS meson decay constants using the results of the eight branches of the analysis carried out in Ref.~\cite{Carrasco:2014cwa} for determining the up, down, strange and charm quark masses. 
The various branches are determined by: ~ i) the choice of the scaling variable, which was taken to be either the Sommer parameter $r_0/a$ \cite{Sommer:1993ce} or the mass of a fictitious PS meson made of two strange-like quarks (or a strange-like and a charm-like quark), $a M_{s^\prime s^\prime}$ (or $a M_{c^\prime s^\prime}$); ~ ii) the fitting procedures, which were based either on Chiral Perturbation Theory (ChPT) or on a polynomial expansion in the light quark mass (for the motivations see the discussion in Section 3.1 of Ref.~\cite{Carrasco:2014cwa}); and ~ iii)  the choice between the methods M1 and M2 (which differ by ${\cal{O}}(a^2)$ effects \cite{Constantinou:2010gr}) used to determine non-perturbatively the values of the mass renormalization constant (RC) $Z_m = 1 / Z_P$.

After correcting for the leading strong isospin-breaking effect due the up- and down-quark mass difference, as determined in Ref.~\cite{Carrasco:2014cwa} at $N_f = 2 + 1 + 1$, the final results obtained for the kaon decay constant and the kaon to pion ratio are
 \bea
    f_{K^+} & = & 154.4 ~ (2.0) \mev ~ , \nn \\
    f_{K^+} / f_{\pi^+} & = & 1.184 ~ (16) ~ , 
    \label{eq:fPScharged}
 \eea
where the errors are the sum in quadrature of the statistical and systematic uncertainties.
In the isospin symmetric limit of QCD we get
 \bea
    f_K & = & 155.0 ~ (1.9) \mev ~ , \nn \\
    f_K / f_\pi & = & 1.188 ~ (15) ~ , \nn \\
    f_D & = & 207.4 ~ (3.8) \mev ~ , \nn \\
    f_{D_s} & = & 247.2 ~ (4.1) \mev ~ , \nn \\
    f_{D_s} / f_D & = & 1.192 ~ (22) ~ , \nn \\
    (f_{D_s} / f_D) / (f_K / f_\pi) & = & 1.003 ~ (14) ~ .
    \label{eq:fPS}
 \eea

\section{Calculation of the kaon decay constant}
\label{sec:fK}

For each ensemble we computed the 2-point PS correlators given by
 \be
    C(t) = \frac{1}{L^3} \sum\limits_{x, z} \left\langle 0 \right| P_5 (x) P_5^\dag (z) \left| 0 \right\rangle \delta_{t, (t_x  - t_z )} ~ ,
    \label{eq:P5P5}
 \ee
where $P_5 (x) = \overline{s}(x) \gamma_5 u(x)$\footnote{Notice that the Wilson parameters of the two valence quarks in any PS meson considered in this work are always chosen to have opposite values. In this way the squared PS meson mass differs from its continuum counterpart only by terms of ${\cal{O}}(a^2 \mu)$ \cite{Frezzotti:2003ni, Frezzotti:2005gi}.}.
At large time distances one has
 \be
    C(t)_{ ~ \overrightarrow{t  \gg a, ~ (T - t) \gg a} ~ } \frac{\mathcal{Z}_K}{2M_K} \left( e^{ - M_K  t}  + e^{ - M_K  (T - t)} \right) ~ ,
    \label{eq:larget}
 \ee
so that the kaon mass and the matrix element $\mathcal{Z}_K = | \langle K | \overline{u} \gamma_5 s | 0 \rangle|^2$ can be extracted from the exponential fit given in the r.h.s.~of Eq.~(\ref{eq:larget}). 
The time intervals $[t_{min}, t_{max}]$ adopted for the fit (\ref{eq:larget}) of the kaon correlation functions can be read off from Table 4 of Ref.~\cite{Carrasco:2014cwa}. 
There they have been determined in a very conservative way by requiring that the changes in the meson masses and decay constants due to a decrease in the value of $t_{min}$ by one or two lattice units are well below the statistical uncertainty.
As far as the charm sector is concerned, we have verified that the contamination of excited states turn out to be practically negligible.
This conclusion can be inferred from the results of Ref.~\cite{Bussone:2014cha}, where the decay constants of charmed pseudoscalar mesons have been computed, on the same lattice ensembles, by using Gaussian smeared operators.

For maximally twisted fermions the value of $\mathcal{Z}_K$ determines the kaon decay constant $f_K$ without the need of the knowledge of any renormalization constant \cite{Frezzotti:2000nk, Frezzotti:2003ni}, namely
 \be
    af_K = a (\mu_\ell + \mu_s) \frac{\sqrt{a^4 \mathcal{Z}_K}}{aM_K ~ \mbox{sinh}(aM_K)} ~ ,
    \label{eq:decayK}
 \ee
where $\mu_\ell$ and $\mu_s$ are the light and strange bare quark masses, respectively.

The statistical accuracy of the correlators (\ref{eq:P5P5}) is significantly improved by using the so-called ``one-end" stochastic method \cite{McNeile:2006bz}, which includes spatial stochastic sources at a single time slice chosen randomly.
Statistical errors on the kaon mass and decay constant are evaluated using the jackknife procedure.

In order to take into account cross-correlations we make use of the eight bootstrap samplings corresponding to the eight analyses of Ref.~\cite{Carrasco:2014cwa} described in Section \ref{sec:intro}.
First, we perform a small interpolation of our lattice data to the value of the strange quark mass, $m_s$, given in Table 13 of Ref.~\cite{Carrasco:2014cwa} for each analysis and corresponding to the average value $m_s = 99.6 (4.3) \mev$\footnote{Throughout this work all the renormalized quark masses are given in the $\rm{\overline{MS}}$ scheme at a renormalization scale of $2 \gev$.}.
Then, we analyze the dependence of the kaon decay constant as a function of the (renormalized) light quark mass $m_\ell \equiv (a \mu_\ell) / (a Z_P)$ and of the lattice spacing $a$, using fitting procedures based either on ChPT or on a polynomial expansion depending on the corresponding analysis of Ref.~\cite{Carrasco:2014cwa}.

The next-to-leading order (NLO) SU(2) ChPT prediction for $f_K$, including discretization and finite size effects, reads as
 \be
    \label{eq:fKCh}
    f_K = P_1 \left( {1 - \frac{3}{4} \xi _\ell \log \xi _\ell  + P_2 \xi _\ell  + P_4 a^2 } \right) \cdot K^{FSE} _f  ~ ,
 \ee
where $\xi_\ell = 2B m_\ell / 16 \pi^2 f^2$, with $B$ and $f$ being the SU(2) low-energy constants (LECs) entering the LO chiral Lagrangian. 
The term proportional to $a^2$ in Eq.~(\ref{eq:fKCh}) accounts for leading discretization effects. 
The factor $K_f^{FSE}$ represents the correction for finite size effects (FSE) in the kaon decay constant, as computed in Ref.~\cite{CDH05} within ChPT.

In the case of the polynomial expansion we adopt the following fit  in $\xi_\ell$:
  \be
     \label{eq:fKPol}
     f_K = P_1^\prime \left( {1 + P_2^\prime \xi _\ell  + P_3^\prime \xi _\ell^2 + P_4^\prime a^2 } \right) \cdot K^{FSE} _f  ~ .
 \ee

\begin{figure}[htb!]

\centering

\scalebox{0.255}{\includegraphics{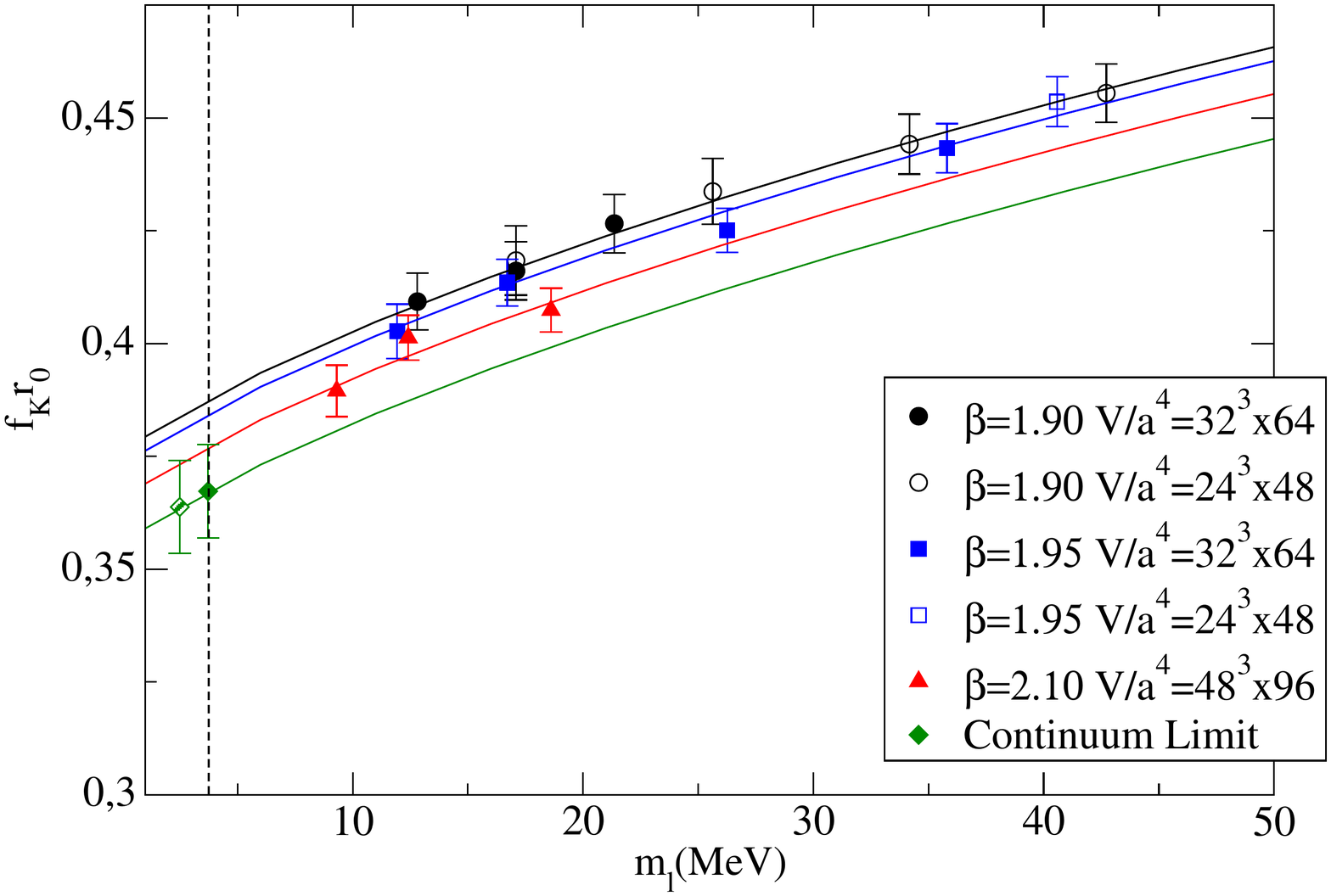}}
\scalebox{0.255}{\includegraphics{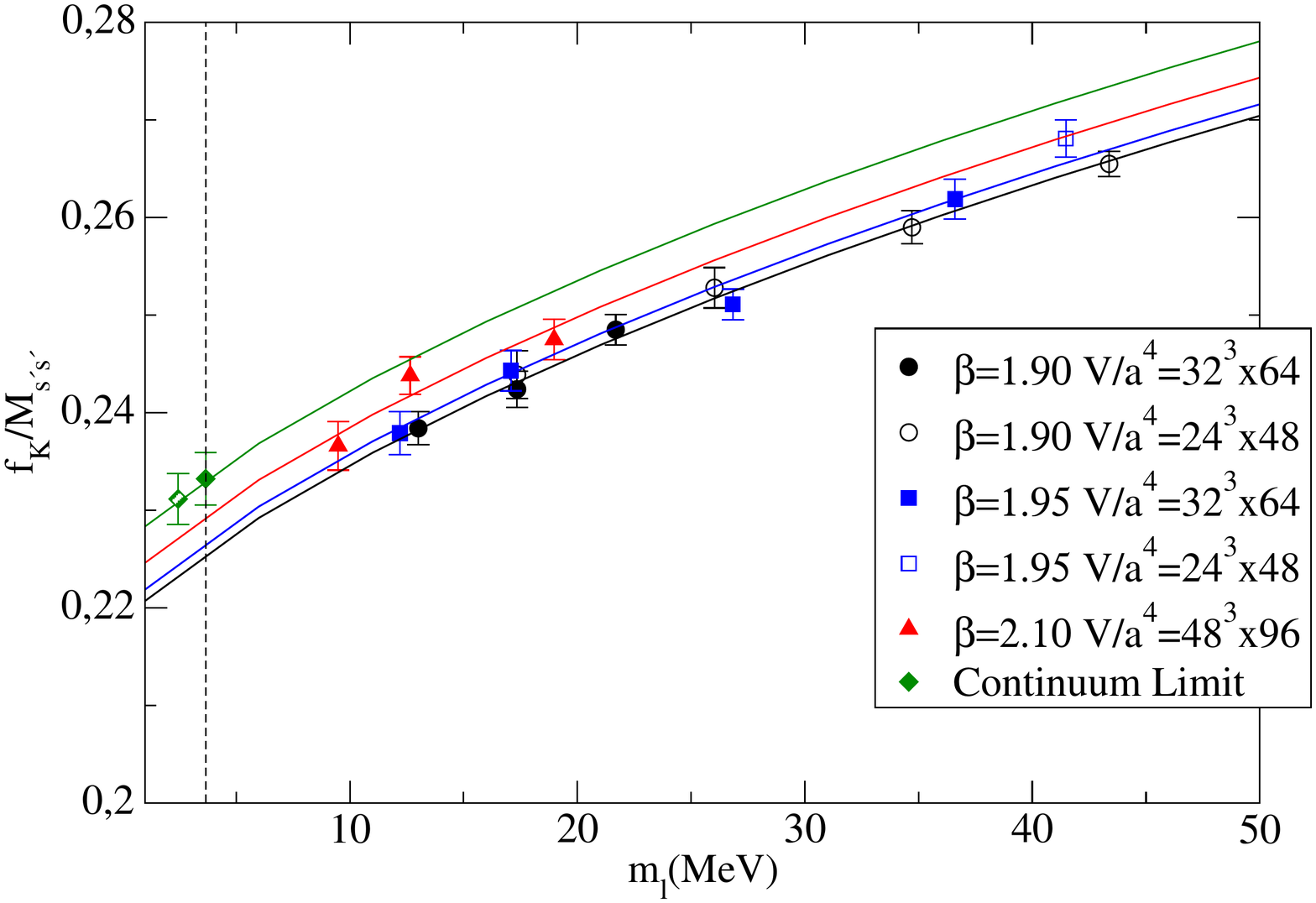}}
\caption{\it Chiral and continuum extrapolation of $f_K r_0$  (left) and $f_K / M_{s^\prime s^\prime}$ (right) based on the NLO ChPT fit of Eq.~(\ref{eq:fKCh}). Lattice data have been corrected for FSE following Ref.~\cite{CDH05}. The green diamond represents the continuum limit evaluated at the average up/down quark mass $m_{ud} = 3.70 (17) \mev$, while the open diamond corresponds to the up-quark mass $m_u = 2.36 (24) \mev$ \cite{Carrasco:2014cwa}.}
\label{fig:fkchir}

\scalebox{0.255}{\includegraphics{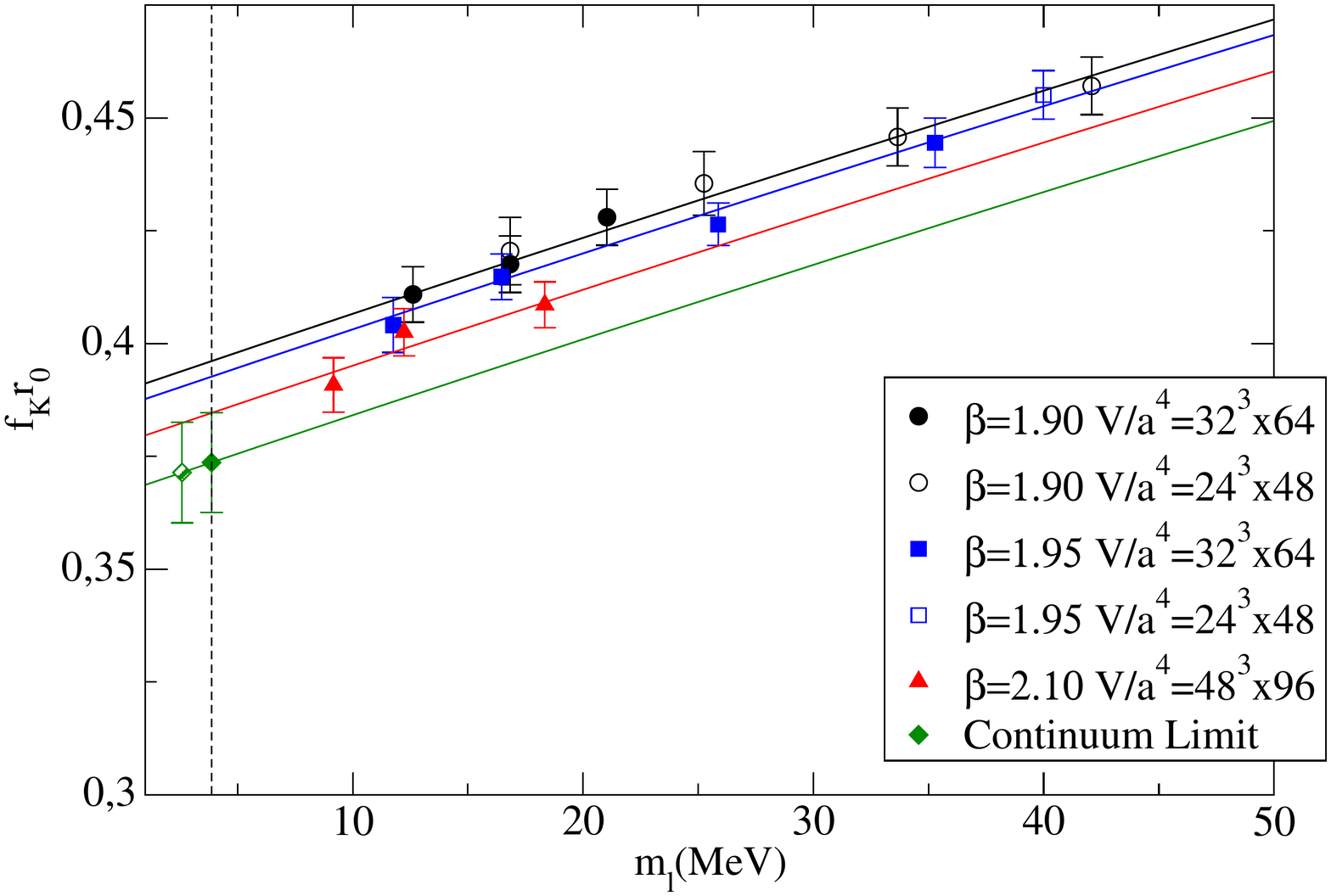}}
\scalebox{0.255}{\includegraphics{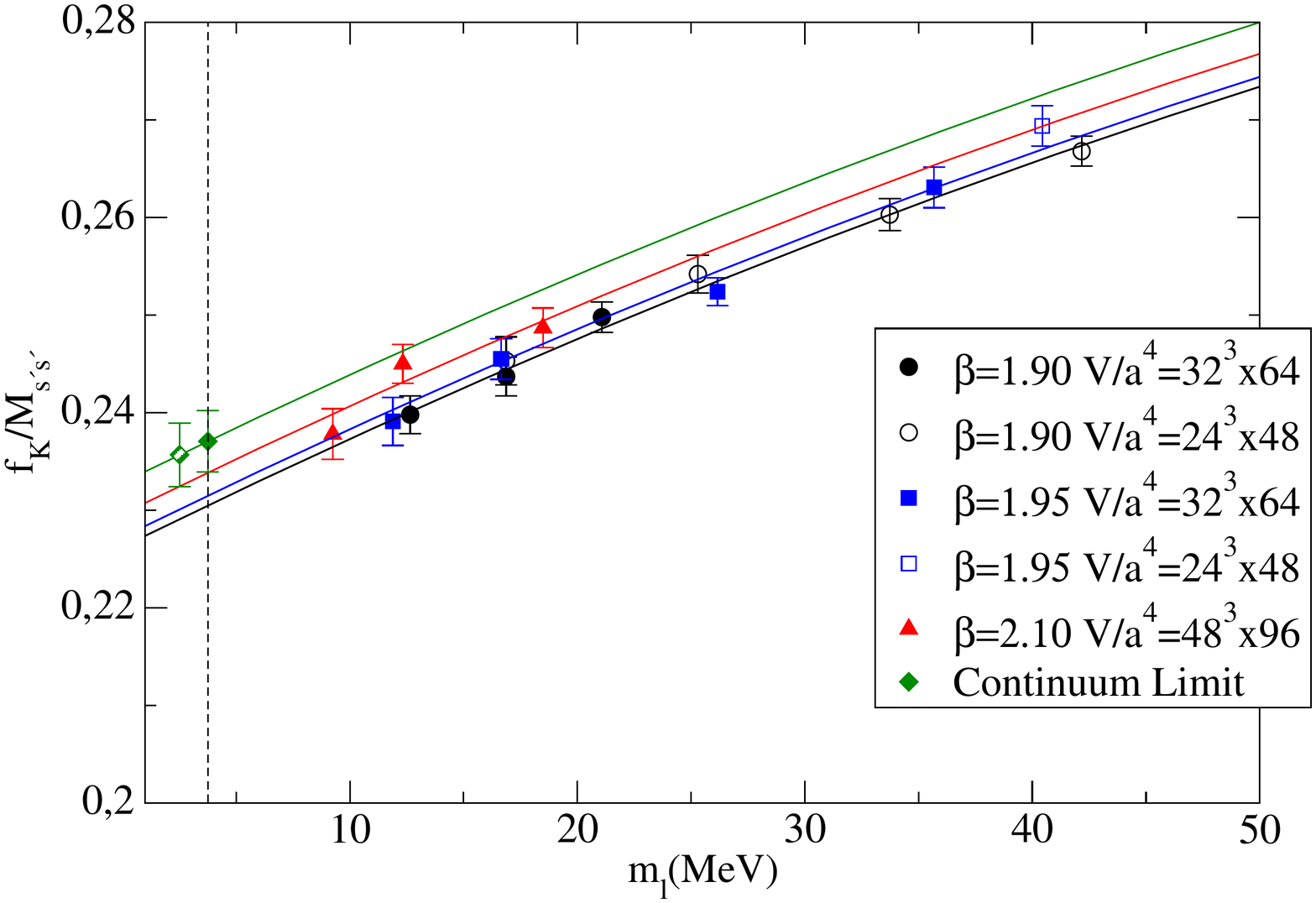}}
\caption{\it The same as in Fig.~\ref{fig:fkchir}, but using for the chiral and continuum extrapolation the polynomial fit of Eq.~(\ref{eq:fKPol}).}
\label{fig:fkpol}

\end{figure}

The (combined) chiral and continuum extrapolation of $f_K $ is shown in Figs.~\ref{fig:fkchir}-\ref{fig:fkpol} in units of either $r_0$ or the mass $M_{s^\prime s^\prime}$ of the fictitious PS meson made of two strange-like valence quarks with mass fixed at $r_0 m_{s^\prime} = 0.22$.
The impact of discretization effects using $r_0$ as the scaling variable is at the level of $\simeq 3 \%$\footnote{The impact of discretization effects is quantified by the spread between the data at the finest lattice spacing and the continuum limiting curve.}, while the use of $M_{s^\prime s^\prime}$, which is affected by cutoff effects similar to the ones of the K-meson mass (without introducing however any significant dependence on the light quark mass), reduces the lattice artefacts down to $\simeq - 1.5 \%$.

Notice in Figs.~\ref{fig:fkchir}-\ref{fig:fkpol} that after taking the continuum limit the kaon decay constant has been extrapolated to two different values of the light quark mass, namely the isospin symmetric, average up/down quark mass $m_{ud} = 3.70 (17) \mev$ and the up-quark mass $m_u = 2.36 (24) \mev$ found in \cite{Carrasco:2014cwa}.
In Section \ref{sec:IB} we will make use of the two extrapolated values of $f_K$ to determine the leading QCD isospin breaking effect due to the mass difference $(m_d - m_u)$ and to provide our result for $f_{K^+}$.

Since the quality of the chiral/continuum fits is quite similar for the various analyses, the corresponding results are combined assuming the same weight for each of them, namely the central value $\overline{x}$ and the variance $\sigma^2$ for an observable $x$ are estimated as (see Ref.~\cite{Carrasco:2014cwa})
\bea
    \overline{x} & = & \frac{1}{N} \sum_{i=1}^N x_i ~ , \nn \\ 
    \sigma^2 & = & \frac{1}{N} \sum_{i=1}^N \sigma_i^2  + \frac{1}{N} \sum_{i=1}^N (x_i  - \overline{x})^2 ~ ,
    \label{eq:combineresults}
 \eea
where $x_i$ and $\sigma_i^2$ are the central value and the variance of the $i$-th analysis and $N$ is the number of analyses ($N = 8$ in our case).
The second term in the r.h.s.~of Eq.~(\ref{eq:combineresults}), coming from the spread among the results of the different analyses, corresponds to a systematic error which accounts for the uncertainties coming from the chiral extrapolation, the cutoff effects and the determination of $Z_P$. 
Finally we add in quadrature to Eq.~(\ref{eq:combineresults}) the systematic uncertainties associated to the calculation of the FSE.

In order to separate the various sources of systematic uncertainties we split the contribution coming from the second term in the r.h.s.~of Eq.~(\ref{eq:combineresults}) into those related to the differences of the results obtained using $r_0$ or $M_{s^\prime s^\prime}$ (labelled as Disc), chiral or polynomial fits (labelled as Chiral) and the two methods M1 and M2 for the RCs $Z_P$ (labelled as $Z_P$).
As for the FSE we compare the results obtained by applying the correction with the ones obtained without correcting for FSE. 
Finally, the error on our determination of the strange quark mass represents another source of uncertainty and it has been included in the (stat+fit) error, which includes also the statistical uncertainty and the error associated with the fitting procedure (i.e.~the amplification of the pure statistical error due to the chiral and continuum extrapolation).

In the isospin symmetric limit we get for $f_K$ the value
 \bea
    \label{eq:fk}
    f_K & = & 155.0 ~ (1.4)_{stat + fit} (0.4)_{Chiral} (1.1)_{Disc} (0.1)_{Z_P} (0.4)_{FSE} \mev \nn \\
          & = & 155.0 ~ (1.9) \mev ~ ,
 \eea
which can be compared with the FLAG averages \cite{FLAG}: $f_K = 158.1 (2.5) \mev$ at $N_f = 2$ from Ref.~\cite{Blossier:2009bx} and $f_K = 156.3 (0.9) \mev$ at $N_f = 2 + 1$ from Refs.~\cite{Follana:2007uv, Bazavov:2010hj, Arthur:2012opa}.
Dividing the result (\ref{eq:fk}) by the experimental value of the pion decay constant, $f_{\pi^+} = 130.41 (20)\mev$ \cite{PDG}, which has been used as input to set the lattice scale \cite{Carrasco:2014cwa}, we get for the ratio $f_K / f_\pi$ the value
 \bea
    \label{eq:fk/fpiresult}
    f_K / f_\pi  & = & 1.188 ~ (11)_{stat + fit} (4)_{Chiral} (9)_{Disc} (1)_{Z_P} (4)_{FSE} (2)_{f_{\pi^+}} \nn \\
                     & = & 1.188 ~ (15) ~ .
 \eea
In order to compare with the analysis of Ref.~\cite{Farchioni:2010tb} we ignore discretization effects and limit ourselves to the gauge ensembles at the two finest lattice spacings corresponding to $\beta=1.95$ and $2.10$.
In this way the value for $f_K / f_\pi$ turns out to be larger by $\simeq 2.5 \%$ with respect to Eq.~(\ref{eq:fk/fpiresult}), getting very close to the result $f_K / f_\pi = 1.224 (13)$ obtained in Ref.~\cite{Farchioni:2010tb}.

\subsection{Mistuning of the strange and charm sea quark masses}
\label{sec:mistuning}

In Ref.~\cite{Carrasco:2014cwa} the strange and charm sea quark masses corresponding to the input bare masses adopted for generating the ETM gauge ensembles at the three values of the lattice spacing, have been determined by comparing data obtained using the OS and the unitary setups for the valence quarks.
For the strange sea quark mass $m_s^{sea}$ we got the values $m_s^{sea} = \{ 99.2 ~ (3.5), ~ 88.3 ~ (3.8),  ~ 106.4 ~ (4.6) \} ~ \mev$  at $\beta = \{1.90, ~ 1.95, ~ 2.10 \}$, which differ from the determination of the strange quark mass, $m_s = 99.6 (4.3)$ MeV, by $\approx 10 \%$ at most, with the largest difference occurring at $\beta = 1.95$.

To estimate the effect of the mistuning of the strange sea quark mass on $f_K$ we use the partially quenched SU(3) ChPT predictions at NLO developed in Refs.~\cite{Bernard:1992mk}-\cite{Bijnens:2006jv} for arbitrary values of sea and valence quark masses, namely
 \bea
      \Delta f_K & \equiv & f_K(m_\ell, m_s; m_s^{sea}) - f_K(m_\ell, m_s; m_s) \nonumber \\  \nonumber 
                       & = & \frac{2}{f_0} \left\{ 4 L_4^r(\mu) \left( \chi_s^{sea} - \chi_s \right) - \frac{1}{12} \overline{A}(\chi_\eta^{sea}) 
                                 \frac{(\chi_s - \chi_\ell)^2 (\chi_\eta^{sea} - \chi_s^{sea})} {(\chi_\eta^{sea} - \chi_s)^2 (\chi_\eta^{sea} - \chi_\ell)} - 
                                 \frac{3}{8} \overline{A}(\chi_\eta) \right. \nonumber \\                                      
                       & - & \left. \frac{1}{12} \overline{A}(\chi_s) \frac{(\chi_s^{sea} - \chi_\ell ) (\chi_s - \chi_\eta^{sea}) - 
                                (\chi_s - \chi_\ell) (\chi_s - \chi_s^{sea})}{(\chi_s - \chi_\eta^{sea})^2} + \frac{1}{4} \overline{A}(\chi_s) \right. \nonumber \\
                       & + & \left. \frac{1}{4} \left[ \overline{A}\left( \frac{\chi_\ell + \chi_s^{sea}}{2} \right) - \overline{A}\left( \frac{\chi_\ell + \chi_s}{2} \right) +
                                 \overline{A}\left( \frac{\chi_s + \chi_s^{sea}}{2} \right) - \overline{A}(\chi_s) \right] \right. \nonumber \\
                       & - & \left. \frac{1}{12} \frac{\partial{\overline{A}(\chi_s)}}{\partial{\chi_s}} \frac{(\chi_s - \chi_\ell) (\chi_s - \chi_s^{sea})} {\chi_s - \chi_\eta^{sea}} 
                                \right\} ~ ,                                  
      \label{eq:fK_msea}
 \eea
where
 \bea
      \chi_\ell & \equiv & 2 B_0 m_\ell ~ , ~ \qquad \qquad  \chi_s \equiv 2 B_0 m_s ~ , \qquad \qquad \chi_\eta \equiv \frac{1}{3} \left( \chi_\ell + 2 \chi_s \right) ~ , 
      \nonumber \\
      \chi_s^{sea} & \equiv & 2 B_0 m_s^{sea} ~ , \qquad \quad \chi_\eta^{sea} \equiv \frac{1}{3} \left( \chi_\ell + 2 \chi_s^{sea} \right) ~  , \nonumber \\      
      \overline{A}(\chi) & \equiv & - \frac{\chi}{16 \pi^2} ~ \mbox{log}\left( \frac{\chi}{\mu^2} \right)
      \label{eq:defs}
 \eea
and $B_0$ and $f_0$ are the SU(3) LECs at LO, while $L_4^r(\mu)$ is a NLO LEC evaluated at the renormalization scale $\mu$.
Taking the values $B_0 / f_0 = 19 ~ (2) $ and $L_4^r(\mu) = 0.04 ~ (14) \cdot 10^{-3}$ at $\mu = M_\rho = 0.770$ GeV from Ref.~\cite{FLAG}, the correction (\ref{eq:fK_msea}) is found to be below the $0.4 \%$ level at our simulated light-quark masses and decreases toward the physical point.
We have checked that by applying the correction (\ref{eq:fK_msea}) to the lattice data the changes observed in the predictions for $f_K$ at the physical point are smaller than $0.3 \mev$.

In a similar way the charm sea quark mass $m_c^{sea}$ has been determined in Ref.~\cite{Carrasco:2014cwa}, obtaining the values $m_c^{sea} = \{ 1.21 (5), ~ 1.21 (5),  ~ 1.38 (4) \} \gev$ at $\beta = \{1.90, ~ 1.95, ~ 2.10 \}$, which should be compared with the determination of the charm quark mass $m_c = 1.176 ~ (39) \gev$.
It follows that, while there is a good agreement at $\beta = 1.90$ and $1.95$, a $\approx 18 \%$ mistuning is present at $\beta= 2.10$.
Since scaling distortions are not visible in our data, we expect that in the continuum limit the mistuning of the charm sea quark mass has a negligible effect compared to the one of the strange sea quark and, therefore, it does not affect our determination of decay constants in a significant way.

\subsection{Isospin breaking effect on the kaon decay constant}
\label{sec:IB}

In this Section we provide an estimate of the isospin breaking (IB) effects on the charged kaon decay constant $f_{K^+}$.
As is known, IB effects are generated by the up and down quark electric charges and by the up and down quark mass difference.
While in the case of hadron masses both QED and QCD IB effects have been determined using a variety of approaches on the lattice, the situation for the decay constant is completely different.
Indeed it is not even possible to give a physical definition to the decay constant in the presence of the QED interaction \cite{Gasser:2010wz}, because of well-known infrared divergencies affecting the calculation of, e.g., the $K_{\ell 2}$ decay rate.
Therefore QED effects on the decay rate of a charged pseudoscalar meson are till now accounted for by relying on ChPT and model-dependent approximations\footnote{A new, promising approach for a lattice determination of QED corrections to generic hadronic processes has recently proposed in Ref.~\cite{Carrasco:2015xwa}}.

In what follows we limit ourselves to the IB effect on $f_{K^+}$ due to the up and down quark mass difference in pure QCD, i.e.~switching off the QED interaction.

Let us consider the decay constant $f_{K^+}$ as a function of the sea $u$- and $d$-quark masses, $m_u^{sea}$ and $m_d^{sea}$, and of the valence $u$-quark mass, $m_u^{val}$ and omit for the sake of simplicity to indicate the dependence on the strange and charm quark masses.
At leading order in the mass differences $(m_u^{sea} - m_{ud})$, $(m_d^{sea} - m_{ud})$ and $(m_u^{val} - m_{ud})$, where $m_{ud}$ is the isospin symmetric, average up/down quark mass, one has
 \bea
      f_{K^+} & = & f_K(m_u^{sea}, m_d^{sea}; m_u^{val}) = f_K(m_{ud}, m_{ud}; m_{ud}) + \left[ \frac{\partial f_K}{\partial m_u^{sea}} \right]_{m_{ud}} 
                            (m_u^{sea} - m_{ud}) \nn \\
                   & + & \left [\frac{\partial f_K}{\partial m_d^{sea}} \right]_{m_{ud}} (m_d^{sea} - m_{ud}) +
                             \left [\frac{\partial f_K}{\partial m_u^{val}} \right]_{m_{ud}} (m_u^{val} - m_{ud}) + ... ~ ,                                                                                      
 \eea
where all the derivatives have to be evaluated at the isospin symmetric point $m_u^{sea} = m_d^{sea} = m_u^{val} = m_{ud}$ and the ellipsis represents terms of higher order.
Since $m_u^{sea} + m_d^{sea} = 2 m_{ud}$ and $\left[ \partial f_K / \partial m_u^{sea} \right]_{m_{ud}} = \left[ \partial f_K / \partial m_d^{sea} \right]_{m_{ud}}$, it follows
 \be
     \label{eq:DeltafKplus}
     f_{K^+} - f_K = \left [\frac{\partial f_K}{\partial m_u^{val}} \right]_{m_{ud}} (m_u^{val} - m_{ud}) + ... ~ ,
 \ee
which means that the leading IB correction to $f_K$ can be obtained from the partial derivative of the decay constant with respect to the valence light-quark mass.

The IB slope $ \left [\partial f_K / \partial m_u^{val} \right]_{m_{ud}}$ can be determined with high precision using the method of Refs.~\cite{deDivitiis:2011eh, deDivitiis:2013xla}, which is based on the insertion of the isovector scalar density in the correlators of the isospin symmetric theory.
This calculation is in progress and will be reported elsewhere.

For the time being we derive an estimate of the partial derivative (\ref{eq:DeltafKplus}) following two methods: ~ i) by adopting the partially quenched SU(3) ChPT developed in Refs.~\cite{Bernard:1992mk}-\cite{Bijnens:2006jv}, and ~ ii) by studying numerically the dependence of the decay constant $f_K$ on the light-quark mass.

The first method totally relies on the partially quenched SU(3) ChPT, which predicts at NLO \cite{Bernard:1992mk}-\cite{Bijnens:2006jv}
 \bea
     \left [\frac{\partial f_K}{\partial m_u^{val}} \right]_{m_{ud}} & = & \frac{4 B_0}{f_0} \left\{ 2 L_5^r(\mu) - \frac{1}{128 \pi^2} 
                        \left[ 1 + 2 \mbox{log}\left( 2 B_0 \frac{m_{ud}}{\mu^2} \right) + \right. \right. \nn \\ 
                        & + & \left. \left. \mbox{log}\left( B_0 \frac{m_s + m_{ud}}{\mu^2} \right) + \mbox{log}\left( \frac{2 m_s + m_{ud}}{3 m_{ud}} \right) \right] \right\}
 \eea
Using the values $B_0 / f_0 = 19 ~ (2) $ and $L_5^r(\mu) = 0.84 ~ (38) \cdot 10^{-3}$ at $\mu = M_\rho = 0.770$ GeV from Ref.~\cite{FLAG}, as well as the values of $m_{ud}$ and $m_s$ determined in Ref.~\cite{Carrasco:2014cwa}, the partial derivative of $f_K$ with respect to the valence light-quark mass is estimated to be equal to $0.37 (7)$ in the $\rm{\overline{MS}}(2 \gev)$ scheme.
This leads to
 \be
     \label{eq:fKplus_1}
      f_{K^+} - f_K = -0.49 ~ (13) \mev ~ ,
  \ee
where the error does not include any estimate of the impact of ChPT orders higher than the NLO one.

In the second method we want to use our non-perturbative results for $f_K$ at $m_\ell = m_u$ and at $m_\ell = m_{ud}$, which have been presented in Figs.~\ref{fig:fkchir}-\ref{fig:fkpol}.
In our simulations, however, the sea and valence light-quark masses are taken to be degenerate and therefore the difference between the two results for $f_K$ at $m_\ell = m_u$ and at $m_\ell = m_{ud}$ does not provide an estimate for $(f_{K^+} - f_K)$. 
Rather we have
 \be
      \label{eq:fKplus}
      f_{K^+} - f_K = f_K(m_u, m_u; m_u) - f_K(m_{ud}, m_{ud}; m_{ud}) + \Delta_{f_K} (m_u - m_{ud}) + ... ~ ,
 \ee
where the ellipsis stands for higher order terms and the correction $\Delta_{f_K}$ is given by
 \bea
     \label{eq:DeltafK}
     \Delta_{f_K} & = & \left[ \frac{\partial f_K(m_\ell^{sea}, m_\ell^{sea}; m_\ell^{val})}{\partial m_\ell^{val}} - 
                                   \frac{d f_K(m_\ell, m_\ell; m_\ell)}{d m_\ell} \right]_{m_\ell^{val} = m_\ell^{sea} = m_\ell = m_{ud}} \nn \\
                         & = & - \left[ \frac{\partial f_K(m_\ell^{sea}, m_\ell^{sea}; m_\ell^{val})}{\partial m_\ell^{sea}} \right]_{m_\ell^{val} = m_\ell^{sea} = m_{ud}} ~ . 
 \eea
We estimate the derivative of the decay constant with respect to the sea light-quark mass using the partially quenched SU(3) ChPT at NLO  \cite{Bernard:1992mk}-\cite{Bijnens:2006jv}, which yields 
 \bea
    \label{eq:DeltafKSU3}
    \Delta_{f_K} & = & \frac{4 B_0}{f_0} \left\{ - 8 L_4^r(\mu) + \frac{1}{64 \pi^2} \left[ 3 + \mbox{log}\left( 2 B_0 \frac{m_{ud}}{\mu^2} \right) +
                                  \mbox{log}\left( B_0 \frac{m_s + m_{ud}}{\mu^2} \right) \right. \right. \nn \\ 
                        & - & \left. \left. \frac{1}{2} \frac{m_s + 2 m_{ud}}{m_s - m_{ud}} \mbox{log}\left( \frac{2 m_s + m_{ud}}{3 m_{ud}} \right) \right] \right\}
 \eea
Using the values $B_0 / f_0 = 19 ~ (2) $ and $L_4^r(\mu) = 0.04 ~ (14) \cdot 10^{-3}$ at $\mu = M_\rho = 0.770$ GeV, the derivative $\Delta_{f_K}$ in the $\rm{\overline{MS}}(2 \gev)$ scheme is estimated to be equal to $\Delta_{f_K} = -0.38 (10)$, which leads to $\Delta_{f_K} (m_u - m_{ud}) = 0.51 (17) \mev$.

From our lattice data (see Figs.~\ref{fig:fkchir}-\ref{fig:fkpol}) we find $f_K(m_u, m_u; m_u) - f_K(m_{ud}, m_{ud}; m_{ud}) = -1.25 (31)$ MeV.
Therefore from Eq.~(\ref{eq:fKplus}) we get the estimate
 \be
     \label{eq:fKplus_2}
      f_{K^+} - f_K = -0.74 ~ (35) \mev ~ ,
  \ee
which is consistent with the estimate of the direct method (\ref{eq:fKplus_1}) within the errors.

Therefore we average the two determinations (\ref{eq:fKplus_1}) and (\ref{eq:fKplus_2}) obtaining our final result
 \be
     \label{eq:fKplusresult}
      f_{K^+} - f_K = -0.62 ~ (29) \mev ~ .
  \ee
Using Eq.~(\ref{eq:fk}) we get 
 \be
    \label{eq:IBslope}
    \frac{f_{K^+} - f_K}{f_K} = -0.0040 ~ (19) ~ ,
\ee
which is quite close to the more precise result $(f_{K^+} - f_K) / f_K = -0.0040 (4)$ obtained with $N_f = 2$ in Ref.~\cite{deDivitiis:2013xla} using a dedicated approach.

Thus, for $f_{K^+}$ we obtain the value
 \bea
    \label{eq:fk+result}
    f_{K^ +} & = & 154.4 ~ (1.5)_{stat + fit} (0.4)_{Chiral} (1.1)_{Disc} (0.1)_{Z_P} (0.4)_{FSE} (0.3)_{(f_{K^+} - f_K)} \mev \nn \\
                  & = & 154.4 ~ (2.0) \mev
 \eea
and, upon dividing Eq.~(\ref{eq:fk+result}) by the experimental value of the pion decay constant, we get
 \bea
    \label{eq:fk+/fpi+results}
    f_{K^+ } / f_{\pi^+} & = & 1.184 ~ (12)_{stat + fit} (3)_{Chiral} (9)_{Disc} (1)_{Z_P} (3)_{FSE} (2)_{f_{\pi^+}} (3)_{(f_{K^+} - f_K)} \nn \\
                                  & = & 1.184 ~ (16) ~ .
 \eea
Our result (\ref{eq:fk+/fpi+results}) can be compared with the FLAG averages \cite{FLAG}: $f_{K^+} /f_{\pi^+} = 1.205 (18)$ at $N_f = 2$ from Refs.~\cite{Blossier:2009bx, deDivitiis:2011eh}, $f_{K^+} /f_{\pi^+} = 1.192 (5)$ at $N_f = 2 + 1$ from Refs.~\cite{Follana:2007uv,  Bazavov:2010hj, Arthur:2012opa, Durr:2010hr} and $f_{K^+} /f_{\pi^+} = 1.194 (5)$ at $N_f = 2 + 1 + 1$ from Refs.~\cite{Bazavov:2013vwa, Dowdall:2013rya}.

\subsection{Determination of $|V_{us}|$}
\label{sec:Vus}

Precision experimental data available for kaon and pion leptonic decays can determine very accurately the ratio $|V_{us} / V_{ud}|$ $f_{K^+} / f_{\pi^+}$, giving $|V_{us} / V_{ud}|$ $f_{K^+} / f_{\pi^+} = 0.2758 (5)$ \cite{Antonelli:2010yf}.
At the same time the determination of $|V_{ud}|$ from superallowed nuclear $\beta$ decays has become remarkably precise: $|V_{ud}| = 0.97425 (22)$ \cite{Hardy:2008gy}.

Therefore, using our result (\ref{eq:fk+/fpi+results}) one obtains
 \be
      |V_{us}| = 0.2269 ~ (4)_{\rm exp} (29)_{f_{K^+} / f_{\pi^+}} = 0.2269 ~ (29) ~ ,
      \label{eq:Vus}
 \ee
where the first error comes from the experimental uncertainties, while the second is due to the uncertainty on $f_{K^+} / f_{\pi^+}$.

Since the CKM matrix is unitary in the Standard Model, the elements of the first row should obey the constraint
 \be
      |V_u|^2 = |V_{ud}|^2 + |V_{us}|^2 + |V_{ub}|^2 = 1 ~ .
      \label{eq:Vu2}
 \ee
The contribution from $|V_{ub}|$ is very tiny, being $|V_{ub}| = 4.13 (49) \cdot 10^{-3}$ \cite{PDG}.
Using our result (\ref{eq:Vus}) one gets
 \be
      |V_u|^2 = 1.0007 ~ (5)_{\rm exp} (13)_{f_{K^+} / f_{\pi^+}} = 1.0007 ~ (14) ~ ,
      \label{eq:Vu2}
 \ee
which confirms the first-row CKM unitarity at the permille level.

\section{Calculation of $f_D$, $f_{D_s}$ and $f_{D_s} / f_D$}
\label{sec:fD}

In this Section we present our determinations of the decay constants $f_D$ and $f_{D_s}$, as well as of the ratio $f_{D_s} / f_D$.
Our analysis is based on the study of the quark mass dependence of two dimensionless ratios, namely $f_{D_s} / M_{D_s}$ and $(f_{D_s} / f_D) / (f_K / f_\pi)$.
Our choice is motivated by the following observations: ~ i) the ratio $f_{D_s} / M_{D_s}$ is affected by smaller discretization effects with respect to other choices like $f_{D_s} r_0$ or $f_{D_s} \sqrt{M_{D_s}} r_0^{3/2}$ (see also Ref.~\cite{Dimopoulos:2013qfa}); ~ ii) the double ratio $(f_{D_s} / f_D) / (f_K / f_\pi)$ exhibits a very mild dependence on the light quark mass  \cite{Becirevic:2002mh} at variance with the ratio $f_{D_s} / f_D$.

For each bootstrap event we perform a small interpolation of the lattice data for $f_{D_s} / M_{D_s}$ and $(f_{D_s} / f_D) / (f_K / f_\pi)$ to the strange and charm quark masses determined in Ref.~\cite{Carrasco:2014cwa}.
The dependences of $f_{D_s} / M_{D_s}$ on the light-quark mass $m_\ell$ and on the lattice spacing turn out to be well described by the simple polynomial expression\footnote{As is known, the Heavy Meson ChPT (HMChPT) predicts no chiral logarithms at NLO for $f_{D_s}$ and $M_{D_s}$. Therefore we have adopted for $f_{D_S} / M_{D_s}$ either a linear ($P_3 = 0$) or a quadratic ($P_3 \neq 0$) expansion in $m_\ell$ [see Eq.~(\ref{eq:fDsMDs})].}
 \be
    \label{eq:fDsMDs}
    f_{D_s} / M_{D_s} = P_1 (1 + P_2 m_\ell  + P_3 m_\ell ^2  + P_4 a^2  ) ~ .
 \ee

\begin{figure}[htb!]
\centering
\scalebox{0.255}{\includegraphics{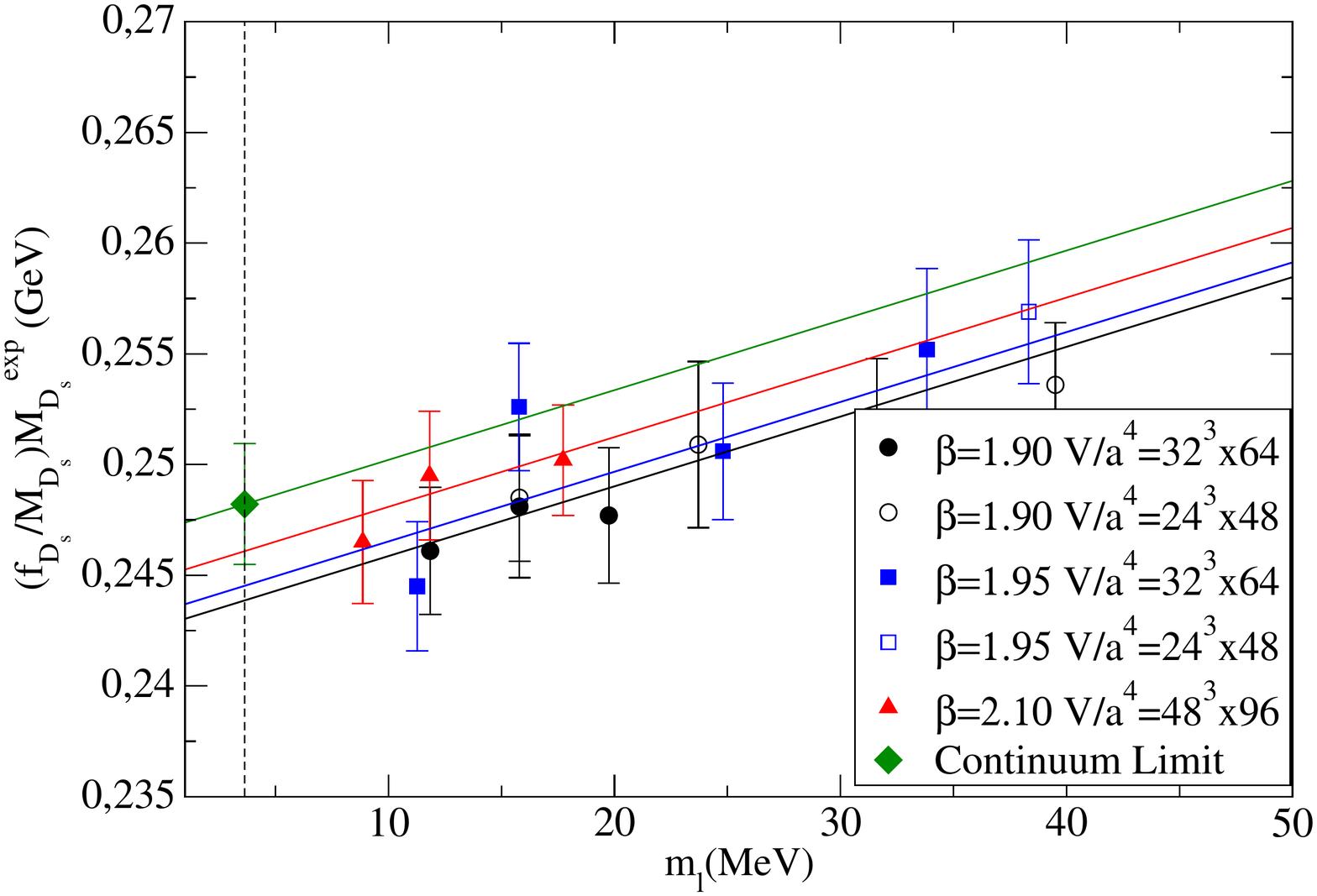}}
\scalebox{0.255}{\includegraphics{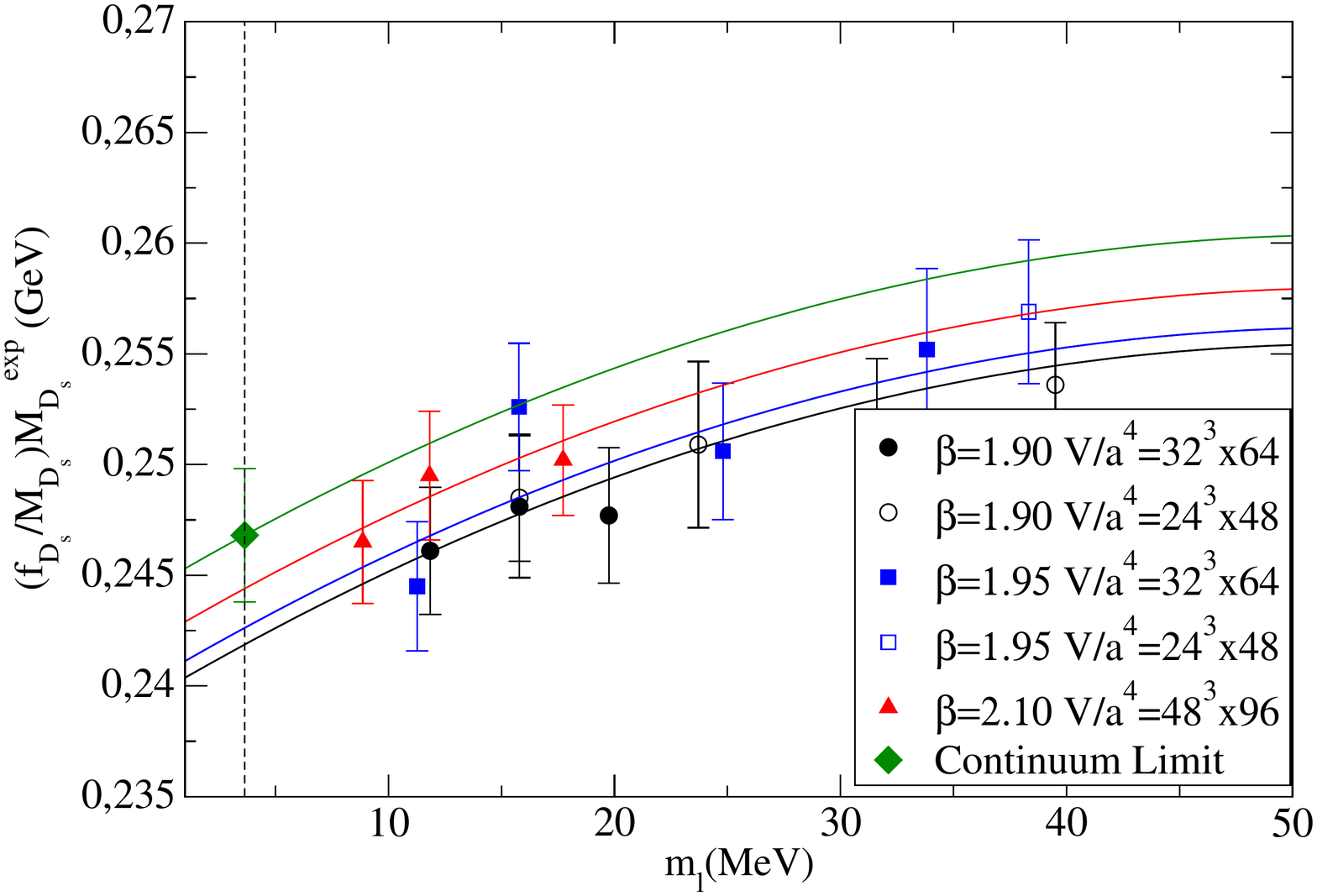}}
\caption{\it Chiral and continuum extrapolation of $(f_{D_s} / M_{D_s}) M_{D_s}^{exp}$ based on Eq.~(\ref{eq:fDsMDs}), assuming $P_3 = 0$ (left) and $P_3 \neq 0$ (right). The diamond represents the continuum limit evaluated at the average up/down quark mass $m_{ud} = 3.70 (17) \mev$ from Ref.~\cite{Carrasco:2014cwa}.}
\label{fig:fDsMDs}
\end{figure}

The chiral and continuum extrapolations of $(f_{D_s} / M_{D_s}) M_{D_s}^{exp}$, obtained according to Eq.~(\ref{eq:fDsMDs}) and using the experimental value $M_{D_s}^{exp} = 1.969 \gev$, are shown in Fig.~\ref{fig:fDsMDs}, where it can be seen that a simple $a^2$-scaling behavior fits nicely our data on $f_{D_s} / M_{D_s}$. 

The systematic uncertainty associated with the chiral extrapolation has been estimated by comparing the results obtained using a linear ($P_3 = 0$) or a quadratic ($P_3 \neq 0$) fit in $m_\ell$, while the one related to discretization effects has been taken from the difference of the results corresponding to the continuum limit and to the finest lattice spacing.
Lattice data corresponding to the same $\beta$ and light quark mass, but different lattice volumes show that FSE are well within the statistical uncertainty. 
Finally, in the (stat+fit) error (quoted below) we have included the errors induced by the uncertainties on the strange and charm quark masses as well as on the input parameters related to the scale setting and to the chiral extrapolation in the light and strange sectors. 

Our final result for $f_{D_s}$ reads
 \bea
    \label{eq:fdsresults}
    f_{D_s} & = & 247.2 ~ (3.9)_{stat + fit} (0.7)_{Chiral} (1.2)_{Disc} (0.3)_{Z_P} \mev \nn \\
                 & = & 247.2 ~ (4.1) \mev 
 \eea
and it can be compared with the FLAG averages \cite{FLAG}: $f_{D_s} = 250 (7) \mev$ at $N_f = 2$ from Ref.~\cite{Carrasco:2013zta} and $f_{D_s} = 248.6 (2.7) \mev$ at $N_f = 2 + 1$ from Refs.~\cite{Davies:2010ip, Bazavov:2011aa}.
Moreover, our result (\ref{eq:fdsresults}) agrees very well with the recent determination $f_{D_s} = 249.0 (0.3)(_{-1.5}^{+1.1}) \mev$ obtained by the FNAL/MILC Collaboration  \cite{Bazavov:2014wgs} with $N_f = 2 + 1 + 1$.

We fit the double ratio $(f_{D_s} / f_D) / (f_K / f_\pi)$ by combining the ChPT predictions for $f_\pi$ and $f_K$ with the HMChPT prediction for $f_{D_s} / f_D$, obtaining 
 \be
    \label{eq:fDsfD}
    \frac{{f_{D_s} / f_D }}{{f_K / f_\pi}} = P_1^\prime \left[ {1 + P_2^\prime m_\ell  + \left( {\frac{9}{4}\hat g^2  - \frac{1}{2}} \right)\xi _\ell \log \xi _\ell } \right] 
                                                             \frac {K^{FSE}_{f_\pi}} {K^{FSE}_{f_K}} ~ ,
 \ee
where for the HMChPT coupling constant $\hat{g}$ we adopt the value $\hat{g} = 0.61 (7)$ \cite{PDG}, which, among the presently available determinations of $\hat{g}$, maximizes the impact of the chiral log in Eq.~(\ref{eq:fDsfD}).
Notice that discretization effects have not been included in Eq.~(\ref{eq:fDsfD}), since within the statistical errors no cutoff dependence is visible in the lattice data (see Fig.~\ref{fig:doubleratio} below). 
As a further check of the impact of discretization effects we perform the fit (\ref{eq:fDsfD}) without including the data at the coarsest lattice spacing (this corresponds roughly to keep half of the data), obtaining the same final result for the double ratio.

In Eq.~(\ref{eq:fDsfD}) we have included the FSE corrections for both $f_\pi$ and $f_K$ taken from Ref.~\cite{Colangelo:2010cu} and Ref.~\cite{CDH05}, respectively.
The former accounts also for the effects of the $\pi^0 - \pi^+$ mass splitting.
In this way the FSE observed in the data at the same light quark mass and lattice spacing but different lattice volumes is correctly reproduced (see Ref.~\cite{Carrasco:2014cwa}).

An alternative fit with no chiral log is performed in order to evaluate the systematic error associated to chiral extrapolation, namely
 \be
     \label{eq:fDsfD_lin}
     \frac{{f_{D_s } /f_D }}{{f_K /f_\pi  }} = \overline{P}_1 \left( {1 + \overline{P}_2 m_\ell} \right) \frac {K^{FSE}_{f_\pi}} {K^{FSE}_{f_K}} ~ .
 \ee
The chiral extrapolations for the double ratio $(f_{D_s} / f_D) / (f_K / f_\pi)$, using either the ChPT (\ref{eq:fDsfD}) or the linear (\ref{eq:fDsfD_lin}) fit, are shown in Fig.~\ref{fig:doubleratio}, where it can be seen clearly that the two fits provide compatible results for all pion masses within the statistical uncertainties.

\begin{figure}[htb!]
\centering
\scalebox{0.45}{\includegraphics{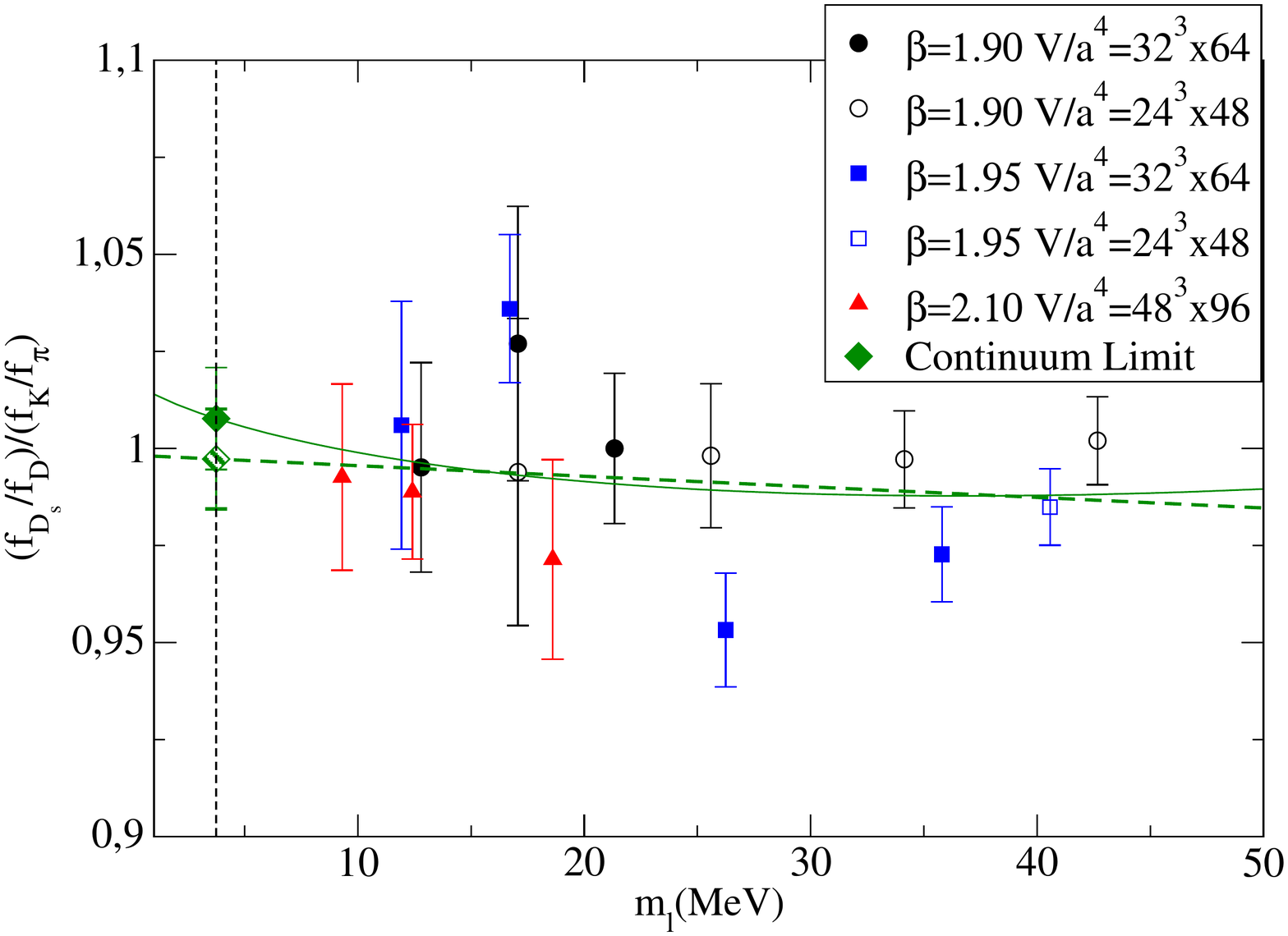}}
\caption{\it Chiral and continuum extrapolation of the double ratio $(f_{D_s} / f_D) / (f_K / f_\pi)$ using both the ChPT fit (\ref{eq:fDsfD}) (solid line) and the polynomial expansion (\ref{eq:fDsfD_lin}) (dashed line) in the light quark mass $m_\ell$.  The full and open diamonds represent the corresponding continuum limit evaluated at the average up/down quark mass $m_{ud}$, respectively. Lattice data have been corrected for FSE using Ref.~\cite{CDH05} for $f_K$ and Ref.~\cite{Colangelo:2010cu} for $f_\pi$.}
\label{fig:doubleratio}
\end{figure}

The most relevant source of systematic errors for the double ratio $(f_{D_s} / f_D) / (f_K / f_\pi)$ is the chiral extrapolation, while for $f_{D_s} / f_D$ also the discretization error coming from $f_K / f_\pi$ is important.
On the other hand, the errors on the strange and charm quark masses, as well as the uncertainty on the RC $Z_P$, contribute negligibly.

Our final results for $(f_{D_s} / f_D) / (f_K / f_\pi)$ and $f_{D_s} / f_D$ are
 \bea
    \label{eq:doubleratioresults}
    \frac{{f_{D_s } /f_D }}{{f_K / f_\pi  }} & = & 1.003 ~ (13)_{stat + fit} (5)_{Chiral} (3)_{FSE} \nn \\
                                                            & = & 1.003 ~ (14) ~ , \\[2mm] 
    \label{eq:fds/fdresults}
    f_{D_s } / f_D  & = & 1.192 ~ (19)_{stat + fit} (8)_{Chiral} (8)_{Disc} (1)_{Z_P} \nn \\
                           & = & 1.192 ~ (22) ~ .
 \eea
The latter one can be compared with the FLAG averages \cite{FLAG}: $f_{D_s } / f_D = 1.20 (2)$ at $N_f = 2$ from Ref.~\cite{Carrasco:2013zta} and $f_{D_s } / f_D = 1.187 (12)$ at $N_f = 2 + 1$ from Refs.~\cite{Bazavov:2011aa, Na:2012iu}.
Notice the remarkable precision for the double ratio (\ref{eq:doubleratioresults}), which means that SU(3) breaking effects in the ratio of PS meson decay constants are the same in the light and charm sectors within a percent accuracy.

Finally we combine our results for $f_{D_s}$ and $f_{D_s} / f_D$ to obtain for $f_D$ the value
 \bea
    \label{eq:fdresults}
    f_D & = & 207.4 ~ (3.7)_{stat + fit} (0.6)_{Chiral} (0.7)_{Disc} (0.1)_{Z_P} \mev \nn \\
           & = & 207.4 ~ (3.8) \mev ~ .
 \eea
The FLAG averages \cite{FLAG} are: $f_D = 212 (8) \mev$ at $N_f = 2$ from Ref.~\cite{Carrasco:2013zta} and $f_D = 209.2 (3.3) \mev$ at $N_f = 2 + 1$ from Refs.~\cite{Bazavov:2011aa, Na:2012iu}.

Our data have been extrapolated to the average up/down quark mass $m_{ud}$ and therefore our results for $f_D$, $f_{D_s}$, $f_{D_s} / f_D$ and $(f_{D_s} / f_D) / (f_K / f_\pi)$ correspond to the isospin symmetric limit of QCD. 

In the case of the D-meson decay constant an estimate of the leading IB effects due to the up- and down-quark mass difference may be obtained in a way similar to the one adopted for the kaon decay constant in Section \ref{sec:IB}.
Using the results of the partially quenched Heavy Meson ChPT (HMChPT) of Refs.~\cite{Sharpe:1995qp, Sharpe} to correct for the derivative of the D-meson decay constant with respect to the sea light-quark mass, we obtain from our lattice data the rough estimate $f_{D^+} - f_D = - 0.4 \pm 0.8 \mev$, which is not inconsistent with the more precise result $f_{D^+} - f_D = 0.47_{-06}^{+25} \mev$ obtained recently in Ref.~\cite{Bazavov:2014wgs}.
However, because of the large error of the above numerical result and of the uncertainty related to the use of an effective field theory valid only in the static limit, we do not provide in this work any estimate for $f_{D^+}$, which is left to a future work, where the method of Refs.~\cite{deDivitiis:2011eh, deDivitiis:2013xla} will be applied.

For the leptonic decay rates of $D$- and $D_s$-mesons we use the latest experimental averages leading to: $f_D |V_{cd}| = 46.06 (1.11) \mev$ and $f_{D_s} |V_{cs}| = 250.66 (4.48) \mev$, as obtained in Ref.~\cite{Rosner:2013ica} by averaging the electron and the muon channels and by including an estimate of structure-dependent Bremsstrahlung effects.
Neglecting other electroweak corrections (see Ref.~\cite{Bazavov:2014wgs} for a first estimate), our results for $f_D$ and $f_{D_s}$ provide the following determinations of the second-row CKM matrix elements: 
 \bea
       \label{eq:Vcdcs}
       |V_{cd}| & = & 0.2221 ~ (53)_{exp} (41)_{f_D} =  0.2221 ~ (67) ~ , \nn \\
       |V_{cs}| & = & 1.014 ~ (18)_{exp} (16)_{f_{D_s}} = 1.014 ~ (24) ~ .
 \eea        

Using $|V_{cb}| = 0.0413 (49)$ \cite{PDG}, the sum of the squares of the second-row CKM elements turns out to be equal to
 \be
      \label{eq:Vc}
      |V_{cd}|^2 + |V_{cs}|^2 + |V_{cb}|^2 = 1.08 ~ (5) ~ ,
 \ee
showing some tension with unitarity.

\section{Conclusions}
\label{sec:conclusions}

We have presented accurate results for the decay constants $f_K$, $f_{K^+}$, $f_D$ and $f_{D_s}$, obtained with $N_f = 2 + 1 + 1$ twisted-mass Wilson fermions.
We have used the gauge configurations produced by the ETMC, which include in the sea, besides two light mass degenerate quarks, also the strange and the charm quarks with masses close to their values in the real world. 
The simulations were based on a unitary setup for the two light mass-degenerate quarks and on a mixed action approach for the strange and charm quarks.
We used data simulated at three different values of the lattice spacing in the range $0.06 \div 0.09$ fm and for pion masses in the range $210 \div 450$ MeV. 

The main results obtained in this paper for the leptonic decay constants of kaon, $D$- and $D_s$-mesons have been collected in Section \ref{sec:intro}, see Eqs.~(\ref{eq:fPScharged})-(\ref{eq:fPS}).

Using the experimental value $|V_{us} / V_{ud}|$ $f_{K^+} / f_{\pi^+} = 0.2758 (5)$ from Ref.~\cite{Antonelli:2010yf} and the updated value $|V_{ud}| = 0.97425 (22)$ from superallowed nuclear $\beta$ decays \cite{Hardy:2008gy}, our result for $f_{K^+} / f_{\pi^+}$ leads to the following determination of the CKM matrix element $|V_{us}|$:
 \be
      |V_{us}| = 0.2269 ~ (29) ~ ,
 \ee
which confirms the unitarity of the first row of the CKM matrix at the permille level, namely
 \be
    |V_{ud}|^2 + |V_{us}|^2 + |V_{ub}|^2 = 1.0007 ~ (14) ~ ,
 \ee
where the contribution from $|V_{ub}|$ is negligible.

Our results for $f_D$ and $f_{D_s}$ combined with the experimental averages of the leptonic decay rates of $D$- and $D_s$-mesons provide the following determinations of the second-row CKM matrix elements
 \bea
       |V_{cd}| & = & 0.2221 ~ (67) ~ , \nn \\
       |V_{cs}| & = & 1.014 ~ (24) ~ ,
 \eea        
which lead to some tension for the unitarity of the second row of the CKM matrix
 \be
      |V_{cd}|^2 + |V_{cs}|^2 + |V_{cb}|^2 = 1.08 ~ (5) ~ ,
 \ee
where the contribution from $|V_{cb}|$ is negligible.

\section*{Acknowledgements}

We thank our colleagues of the ETM Collaboration for fruitful discussions and we are very grateful to S.R.~Sharpe for providing us the explicit expressions for $f_D$ and $f_{D_s}$ in the partially quenched HMChPT with $N_f = 2 +1$.

\noindent We acknowledge the CPU time provided by the PRACE Research Infrastructure under the project PRA027 ``QCD Simulations for Flavor Physics in the Standard Model and Beyond'' on the JUGENE BG/P system at the J\"ulich SuperComputing Center (Germany), and by the agreement between INFN and CINECA under the specific initiative INFN-RM123 on the Fermi system at CINECA (Italy).

\noindent V.~L., S.~S.~and C.~T.~thank MIUR (Italy) for partial support under Contract No. PRIN 2010-2011.

\noindent R.~F.~and G.C.~R.~thank MIUR (Italy) for partial support under Contract No. PRIN 2009-2010.

\end{document}